\documentclass[conference]{IEEEtran}
\IEEEoverridecommandlockouts
\usepackage{cite}
\usepackage{amsmath,amssymb,amsfonts}
\usepackage{algorithmic}
\usepackage{graphicx}
\usepackage{textcomp}
\usepackage{xcolor}

\def\BibTeX{{\rm B\kern-.05em{\sc i\kern-.025em b}\kern-.08em
    T\kern-.1667em\lower.7ex\hbox{E}\kern-.125emX}}
    
\begin{document}

\title{Securing Swarms: Cross-Domain Adaptation for ROS2-based CPS Anomaly Detection\\
\thanks{This work was supported by Clemson University’s Virtual Prototyping of Autonomy Enabled Ground Systems (VIPR-GS), under Cooperative Agreement W56HZV-21-2-0001 with the US Army DEVCOM Ground Vehicle Systems Center (GVSC). DISTRIBUTION STATEMENT A. Approved for public release; distribution is unlimited. OPSEC\#9868
}
}

\author{\IEEEauthorblockN{Julia Boone, Fatemeh Afghah}
\IEEEauthorblockA{Holcombe Department of Electrical and Computer Engineering, Clemson University, Clemson, SC, USA\\
\{jcboone, fafghah\}@clemson.edu}
}

\maketitle

\begin{abstract}
Cyber-physical systems (CPS) are being increasingly utilized for critical applications. CPS combines sensing and computing elements, often having multi-layer designs with networking, computational, and physical interfaces, which provide them with enhanced capabilities for a variety of application scenarios. However, the combination of physical and computational elements also makes CPS more vulnerable to attacks compared to network-only systems, and the resulting impacts of CPS attacks can be substantial. Intelligent intrusion detection systems (IDS) are an effective mechanism by which CPS can be secured, but the majority of current solutions often train and validate on network traffic-only datasets, ignoring the distinct attacks that may occur on other system layers. In order to address this, we develop an adaptable CPS anomaly detection model that can detect attacks within CPS without the need for previously labeled data. To achieve this, we utilize domain adaptation techniques that allow us to transfer known attack knowledge from a network traffic-only environment to a CPS environment. We validate our approach using a state-of-the-art CPS intrusion dataset that combines network, operating system (OS), and Robot Operating System (ROS) data. Through this dataset, we are able to demonstrate the effectiveness of our model across network traffic-only and CPS environments with distinct attack types and its ability to outperform other anomaly detection methods.  
\end{abstract}


\section{Introduction}

Cyber-physical systems (CPS), such as unmanned aerial vehicle (UAV) or unmanned ground vehicle (UGV) swarms, offer substantial benefits for civilian and military applications \cite{uav_use_cases,ugv_use_cases}. CPS systems are unique in containing both computational and physical components, with many CPS also utilizing networked connections to transmit and receive data. Given the strong coupling of hardware (sensors, actuators, etc) with software and network components in CPS, they are capable of influencing and optimizing physical processes across various application domains. As such systems become more pervasive in day-to-day life, existing and new cyber vulnerabilities threaten to disrupt the use of CPS. As CPS are intended to be used to help optimize critical infrastructure and real-world processes, downtime and data leakages from attack CPS can lead to substantial physical and economic damages. Security in CPS is unique in comparison to traditional cybersecurity as the safety, reliability, and real-time capabilities of the system must also be ensured in addition to traditional data confidentiality and integrity concerns.

Robot Operating System (ROS) \cite{quigley2009ros} and ROS2 \cite{ros2} are important in the development of CPS as they provide the frameworks for the development of robotic and automation applications. ROS2, in particular, is equipped with features that are especially relevant for CPS, such as real-time functionality support, Data Distribution Service (DDS)-based security measures (implemented via Secure ROS2 (SROS2) \cite{mayoral-vilcheSROS2}), and multi-platform support. It also removes the single-point-of-failure security concern caused by ROS's Master node, which controlled all communication for ROS-based systems. SROS2 and architectural changes provide critical security infrastructure for ROS2-based CPS systems in military and critical application scenarios, positioning ROS2 to be widely adopted in these domains. Even still, the ability to augment existing security technologies with intelligence is a key advantage of intrusion detection systems and can greatly enhance their overall system security and provide multiple points of verification for these systems.

Anomaly detection methods are crucial for developing intrusion detection systems (IDS) for CPS. The application of machine learning and deep learning for the detection of anomalies and intrusions, particularly in the context of cybersecurity, is a growing field bolstered by recent advancements in ML and DL. Traditional IDS rely on pre-defined rules, but these systems fail when faced with novel or otherwise unknown attacks. ML and DL can both provide IDS with the ability to learn normal network traffic and detect anomalies as deviations from that normal. DL-based approaches tend to perform well with identifying patterns in raw network traffic data while ML-based approaches can utilize previously learned patterns to determine anomalous deviations \cite{Muneer_IDSreview}. The evolving nature of CPS environments necessitates frequent model retraining, making methods reliant on labels resource-intensive to sustain. To address this, many intelligent IDS works focus on the implementation of methods that do not rely on labels to better mimic real-world circumstances in which attack data points are not well defined before system deployment. 

While there is existing work on the detection of attacks in CPS, there is a substantial lack of CPS-specific data intrusion detection work. Many current IDS work focus on network traffic-only systems \cite{boone2025jointreconstructiontripletlossautoencoder,rezakhani2025transferlearningframeworkanomaly}, with network anomaly detection datasets such as NSL-KDD \cite{nsl_kdd} being used for IDS evaluation. CPS IDS trained on network-only data may struggle to detect attacks on other layers of the system, such as ROS2 node crashing attacks \cite{Puccetti2024ROSPaCe}. Additionally, with a growing focus on the integration of Zero Trust (ZT) architectures \cite{nist_zero_trust} into CPS, it is crucial to have an IDS that looks at all layers of a system as opposed to a single layer alone. Thus, evaluating IDS on true CPS data as opposed to network data alone is critically important for the security of these systems.

To bridge this gap and avoid the prohibitive cost of labeling heterogeneous CPS data, we propose a cross-domain anomaly-detection framework that transfers discriminative knowledge learned from a labeled network-traffic environment to an unlabeled multi-layer CPS (network, OS, ROS, etc) environment. The proposed domain adaptation-based anomaly detection approach offers significant benefits for attack detection in CPS, particularly in scenarios where the nature of the traffic,  or attack patterns can vary across domains. The use of domain adaptation improves generalization and robustness, allowing the detection system to adapt to variations in the type of systems evaluated without the need for labels from the target domain. Additionally, domain adaptation helps mitigate the need for large, labeled datasets in the target domain, thus reducing the overhead associated with manual labeling, while enabling the model to detect both known and novel attacks more effectively. 

Our contributions can be summarized by the following:
\begin{itemize}
    \item We present a novel domain adaptation model that uses a combination of contrastive learning, representation learning, clustering, and adversarial training techniques for CPS attack detection. To the best of our knowledge, this is the first method to perform domain adaptation for attack detection from network-only data to multi-layer CPS data. We also develop and utilize a unique multi-flow approach for representation learning in time series data.
    \item Using our method and clustering-based anomaly decisioning, we achieve 98\% attack detection accuracy on our labeled, network-only source data and 87\% accuracy on our unlabeled, multi-layer CPS target data. 
\end{itemize}

\section{Related Work}





\subsection{CPS Intrusion Detection}

Compared to network-focused solutions, there are few works that explore the multiple architectural layers involved in CPS systems that may contribute to attack detection capabilities. \cite{Puccetti2024ROSPaCe} presents ROSPaCe, the first intrusion dataset to provide synchronized data from the network, operating system, and ROS-layers of a CPS system. The authors also provide preliminary results on an unsupervised machine learning algorithm, Isolation Forest, and a supervised machine learning algorithm, XGBoost. \cite{latency_rospace_puccetti} utilizes ROSPaCe to benchmark the latency of five different machine learning algorithms used for anomaly detection.

Crucially, other existing CPS intrusion works often utilize network-only datasets like NSL-KDD \cite{nsl_kdd} or UNSW-NB15 \cite{UNSW-NB15}. \cite{ZoppiMetaLearning} presents a meta-learning based methodology in which an ensemble of unsupervised algorithms are used to perform intrusion detection in CPS, validating on 21 network-only datasets. They find that the use of meta-learning helps reduce attack misclassifications when detecting zero-day/unknown attacks. In \cite{CATILLO_CPSGUARD}, the authors present CPS-GUARD, an intrusion detector for CPS systems that uses a single semi-supervised autoencoder and reconstruction error thresholding to perform intrusion detection. While achieving between 0.915 and 1.000 recall and false positive rates as low as 0.009, this CPS-intended methodology is also validated on network-only data. Evaluating CPS systems using only network data may miss attacks at other layers, such as ROS2 Node Crashing in the ROSPaCe dataset \cite{Puccetti2024ROSPaCe}.

\subsection{Security in ROS-based Systems}

ROS \cite{quigley2009ros} is the primary framework used for the development of robotic applications. Although ROS is well equipped for this development task, it was not designed with inherent security, leaving safety-critical robotics and CPS developed with ROS vulnerable to attack. Security solutions for ROS 1, such as Secure ROS (SROS) \cite{SROS1}, provide some preliminary security features, but these solutions are not well-developed or maintained enough to provide ROS1 adequate security. Additionally, ROS1's centralized design with the ROS Master is insecure as it creates a single point of failure for the entire system. 

Conversely, ROS 2 \cite{ros2} adopts a decentralized design where an abstract middleware interface is implemented. Serialization, transport, and discovery are provided through this middleware interface. By default, ROS2 utilizes Data Distribution Service (DDS) middleware implementations. One key advantage of utilizing DDS middleware implementations for ROS2 is that DDS provides a security specification, DDS-Security, that defines a set of security plugins and plugin implementations. Secure ROS2 (SROS2) \cite{mayoral-vilcheSROS2} is the set of tools and features within ROS2 that are used to enable integration with DDS-Security. However, as with traditional cybersecurity systems, the security of ROS 2 is not inherently guaranteed with SROS2 alone. For instance, \cite{deng_insecureros2} explores several vulnerabilities of ROS2, such as issues with ROS2's access control system or exploits with synchronization. There is limited work, in part due to the limited number of ROS-based intrusion datasets like ROSPaCe, on the application of intrusion detection methods for ROS2 security. \cite{da2025deep} simulates ROS2 DoS attacks via the flooding of ROS2 topic with messages, utilizing four different DL architectures to perform attack detection. Three of these methods are supervised, achieving 97\% accuracy with a multi-layer perceptron architecture, while the only unsupervised methodology, a variational autoencoder (VAE), yields 64\% detection accuracy. Given the potential ubiquitousness of ROS2 in military and critical application scenarios, ROS-targeted IDS systems can be a useful way to enhance system security.

\section{Methodology}

\subsection{Problem Definition}

\begin{figure*}[ht!]
   \centering
    \includegraphics[width=0.90\linewidth]{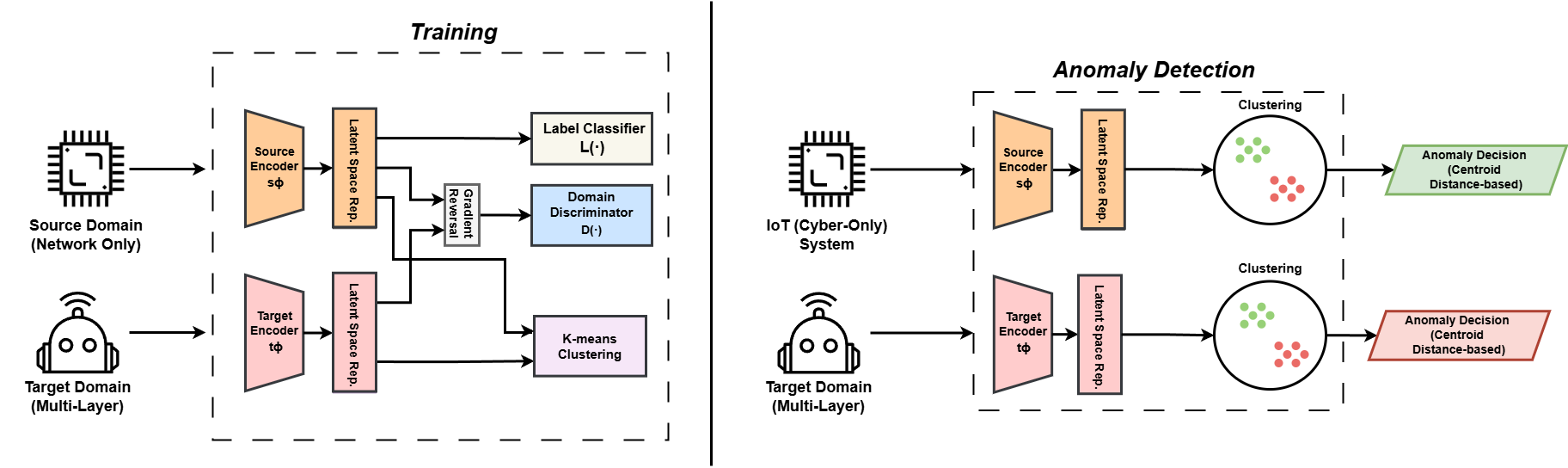}
    \caption{Proposed domain adaptation for CPS-based anomaly detection architecture, training and anomaly detection pipeline. 
    }
    \label{fig:modelarch}
\end{figure*}

In this work, we address the issue of CPS attack detection. We consider a case in which we have access to labeled network traffic-only attack data from a source domain ($D_S$). We want to utilize the labeled source domain's attack information and translate it to a new, unlabeled target domain ($D_T$). We specifically aim to leverage the network traffic-only dataset to detect attacks in a CPS environment that contains data from different modalities (such as OS, ROS, network, etc).

Our goal is to develop a model that transfers discriminative attack knowledge from $D_S$ to $D_T$ by aligning their latent spaces, enabling effective cross-domain detection without requiring labeled data in the target domain.


Our proposed framework for addressing this task is visualized in Figure \ref{fig:modelarch}. In the following sections, we give more details on each component of the model. 

\subsection{Dual Feature Encoders}

In this work, we utilize a dual-encoder architecture to independently transform source and target domain data into lower-dimensional latent representations. The source encoder is denoted as $s_\phi$ and the target encoder is denoted as $t_\phi$. Latent representations are generated in the following manner: 
\begin{equation}
    h_{is} = s_\phi(x_{is}) \quad\text{and}\quad   h_{it} = t_\phi(x_{it})
\end{equation}
where $x_{is}$ is a source domain sample, $x_{it}$ is a target domain sample, $h_{is}$ is the source's latent feature representation, and $h_{it}$ is the target's latent feature representation. For the source encoder, we implement a 1D convolutional neural network (CNN). For the target encoder, we implement a 1D CNN that leverages residual blocks to further improve the feature learning for the target domain, given its lack of structured labels to assist in forming its representations.

\subsection{Sequence Generation and Contrastive Learning}

In our work, we find that the datasets used for both network-only and multi-layer CPS are collected in the format of flows; that is, that each individual sample in the dataset is a collection of aggregated metrics over some duration of time. However, when learning from large datasets collected over an extended period of time, it can be advantageous to sequence these flows. Sequencing allows us to look at extended periods of system behavior to establish better benign behavior pattern baselines and better baselines of how attacks can propagate throughout a system. Sequencing is a common approach in time series forecasting \cite{timeseriesforecasting_benidis} and effectively allows us to model network flows as multivariate time series data. We utilize the same sequence length in the source and target domains to allow representations from the source and target to be used together in the model's label classifier and domain discriminator (detailed in further sections).


To enhance our model's ability to generalize between the source and target domains, we utilize contrastive learning, which has been used with demonstrated success in other domain adaptation tasks \cite{Thota_2021_CVPR}. Our primary goal in utilizing contrastive learning is to learn latent features that are both discriminative between classes and robust to shifts between domains. Here, we employ contrastive learning using a triplet-based approach, where we utilize a triplet margin loss for the actual contrastive loss. The triplets consist of anchor (or original) sequences, positive sequences (sequences that should be closer to the anchor sequences in a latent space due to their similarity), and negative sequences (sequences that should be further away from anchor sequences due to their differences). The generation of these triplets is different between our two domains; we utilize a supervised technique that leverages class labels for the source domain and an unsupervised technique that leverages the temporal proximity of sequences in the target domain.

We denote anchor sequences as $x_{is}$ for the source domain and $x_{it}$ for the target domain. For supervised triplet generation, given a source domain point $x_{is}$ with a corresponding label 0 for benign or 1 for anomaly, we select the negative, $n_{is}$, by randomly sampling from the opposite class.  We select the positive, $p_{is}$, by randomly sampling from the same class. For instance, if $x_{is} \in S_{0}$, the benign set, then the positive sample will be $p_{is} \sim S_{0}$ and negative sample will be $n_{is} \sim S_{1}$, the anomaly set.

\begin{figure}
    \centering
    \includegraphics[width=0.60\linewidth]{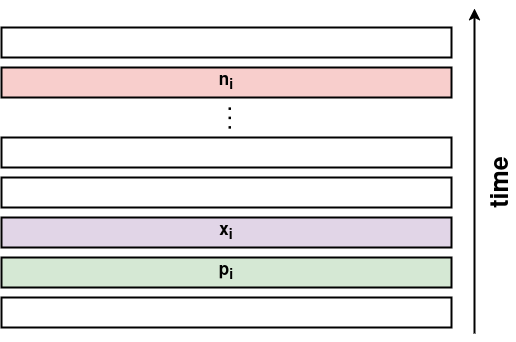}
    \caption{Unsupervised triplet generation for unlabeled target domains.}
    \label{fig:unsupervisedtriplet}
\end{figure}

For unsupervised triplet generation, selecting a positive or negative sample in relation to the anchor sequence is more challenging, as no class labels are available. Here, similar to \cite{NEURIPS2019_53c6de78}, we look at time-based sampling to overcome the lack of labeled data. For positive target samples, $p_{it}$, we consider that the representation of an anchor sequence will be similar to sequences that are close to it in time and select a random sequence that fits this criteria. For the negative target samples, $n_{it}$, we consider that the representation of an anchor sequence will likely not be similar to a sequence that is further away from it in time and select a random sequence that fits this criterion. This process is illustrated in Figure \ref{fig:unsupervisedtriplet}.

In our work, the utilized triplet margin loss is defined as:
\begin{equation}
    \mathcal{L}_{TML} = max\{d(x_i,p_i) - d(x_i,n_i) + m, 0\}
\end{equation}
where  $m$ is the margin value and $d(x_i,y_i)$ is the Euclidean distance. We utilize this loss for both the source and the target domains.

\subsection{Source-domain Label Classifier}

The purpose of the source-domain label classifier, $L(\cdot)$, is to learn the relationship between the source representations, $h_{is}$, generated by the source encoder, $s_{\phi}$, and the attack classifications utilized. In our case, as we consider a binary (attack or no attack) classification scenario, this classifier is designed to guide the latent space into two distinct divisions. We utilize the binary cross-entropy (BCE) loss for our classification loss, denoted as $\mathcal{L}_{class}$.



\subsection{Domain Discriminator}

A key challenge for our model is that we need to align the representations from the source and target domains so that the source attack mapping can be utilized to make distinct classes recognizable during the attack detection task. However, we also need to retain domain-invariant features so that the model can generalize across both domains. To address this, in line with \cite{pmlr-v37-ganin15}, we utilize a domain discriminator $D(\cdot)$ trained on representations from both domains. Before passing these representations to $D(\cdot)$, they are processed with a gradient reversal layer (GRL). The GRL acts as an identity function during forward propagation, but multiplies the incoming gradient from the domain discriminator's loss by a factor of $-\lambda$ during back propagation. The GRL is designed to encourage the feature encoders to produce representations that can "fool" the domain discriminator and maximize its loss, $\mathcal{L}_{D}$. This creates an adversarial training scenario in which $D(\cdot)$ attempts to minimize $\mathcal{L}_{D}$ to accurately differentiate between domains while the feature encoders aim to maximize this loss by promoting feature in-distinguishability. Ideally, the result features will be generalizable across domains while retaining specificity for the attack detection task. 

\begin{table*}[ht]
    \centering
    \caption{Comparison of utilized datasets}
    \begin{tabular}{|c|c|c|c|c|}
        \hline
        \textbf{Dataset} & \textbf{Sample \#} & \textbf{Attacks} & \textbf{Duration} & \textbf{Layers} \\
        \hline
        WUSTL-2021 & 1,194,464 & SQL Injection, DoS, Recon, Backdoor & 53 hours & Network-only \\
        \hline
        ROSPaCe & 30,247,050 & Nmap, ROS2 Recon, Flood-Meta, Flood-Nmap, Node Crash, Reflection & 4 Days & ROS, Network, OS \\
        \hline
    \end{tabular}
    \label{tab:dataset_descriptions}
\end{table*}

\begin{table*}[ht!]
    \centering
    \caption{Accuracy metrics for baseline models \cite{zhao2019pyod} and our proposed method}
    \begin{tabular}{|c|c|c|c|c|c|c|c|c|}
        \hline
         & \multicolumn{4}{|c|}{\textbf{WUSTL-2021}} & \multicolumn{4}{|c|}{\textbf{ROSPaCe}} \\
         \hline
         & \textbf{Accuracy} & \textbf{Precision} & \textbf{Recall} &\textbf{ F-one} & \textbf{Accuracy} & \textbf{Precision} & \textbf{Recall} & \textbf{F-one}\\
         \hline
         \textbf{Ours} & 98.9849 & 0.9886 & 0.9995 & 0.9940 & 87.8753 & 0.8877 & 0.9539 & 0.9164\\
         \hline
         \textbf{DSVDD (benign)} & 92.8948 & 0.7988 & 1.0000 & 0.8881 & 86.6545 & 0.9896 & 0.8635 & 0.9223\\
         \hline
         \textbf{DSVDD (mix)} & 97.2921 & 0.7288 & 0.9985 & 0.8426 & 20.8962 & 0.4468 & 0.0573 & 0.1016\\
         \hline
         \textbf{GMM (benign)} & 92.8945 & 0.7988 & 1.0000 & 0.8881 & 94.6703 & 0.9467 & 1.0000 & 0.9726\\
         \hline
         \textbf{GMM (mix)} & 97.2858 & 0.7284 & 0.9983 & 0.8422 & 20.2642 & 0.4153 & 0.0533 & 0.0945\\
         \hline
         \textbf{LODA (benign)} & 89.3052 & 0.9582 & 0.6492 & 0.7740 & 88.7708 & 0.9937 & 0.8870 & 0.9373 \\
         \hline
        \textbf{LODA (mix)} & 93.8881 & 0.5936 & 0.5001 & 0.5428 & 14.0189 & 0.1025 & 0.0131 & 0.0234\\
         \hline
    \end{tabular}
    \label{tab:baselinespyod_and_ours}
\end{table*}

\subsection{Dunn Index-based Center Loss}

In our work, we also consider that we can enforce stricter regularization on the latent space when we are utilizing the latent representations of the benign and anomaly classes to perform attack detection. In particular, we consider that it is advantageous if the latent representations are in tight, well-formed clusters that do not have significant overlap. While all of the prior loss functions we utilize in this work are beneficial for shaping useful latent representations, they do not explicitly attempt to guide the representations into these clusters for the attack detection task.

To this end, we utilize the Dunn Index \cite{dunn} to formulate an additional loss term. The Dunn Index is the ratio of the minimum of the inter-cluster distances and the maximum of the intra-cluster distances. Here, driving benign and attack samples closer to other samples of the same class, while pushing the two classifications away from one another, is ideal for the formulation of distinct clusters for each classification. The Dunn Index is derived from clustering and quantifies this exact relationship; that is, the distance of distinct clusters from one another and the compactness of like-samples in their individual clusters. The Dunn Index is higher when clusters are tightly formed with larger distances between individual clusters. We define as our loss as the inverse of the Dunn index: 

\begin{equation}
    \mathcal{L}_{Dunn} = \frac{1}{DI_{m}} = \frac{\displaystyle \max_{1 \le k \le m}\Delta_{k}}{\displaystyle \min_{1 \le i < j \le m}
        \delta(C_{i},C_{j})}
\end{equation}

where $\delta(C_{i},C_{j})$ is the inter-cluster distance between clusters and $\Delta_{k}$ is the intra-cluster distances.

With this term, the overall loss terms for our model can be defined as such:
\begin{equation}
\begin{split}
    \mathcal{L}_{adapt} = &\; \lambda_{Dunn} * \mathcal{L}_{Dunn} + \lambda_{TML} * \mathcal{L}_{TML} \\
    & + \lambda_{class} * \mathcal{L}_{class}
\end{split}
\end{equation}

\begin{equation}
    \mathcal{L}_{D} = \lambda_{domain} * \mathcal{L}_{domain}
\end{equation}
where $\lambda_{Dunn}$, $\lambda_{TML}$, $\lambda_{class}$, and $\lambda_{domain}$ are the respective loss weights for each corresponding loss term. Because the domain classifier functions as an adversarial discriminator, it has its own loss function and weight, $\lambda_{domain}$.

\subsection{Clustering-Based Anomaly Decisions}

For anomaly detection, we apply a K-means clustering algorithm to the latent space representations generated by the source and target encoders. Given the binary nature of our problem (attack vs no attack), we configure K-means with K = 2. The K-means algorithm partitions the latent data points into two clusters by minimizing the within-cluster sum of squared Euclidean distances, identifying two distinct centroids, $\mu_1$ and $\mu_2$. For an unlabeled test sample $x_i$, its classification is determined by its proximity to the centroids generated by K-means, where it is assigned to the class of the closest centroid.
\begin{equation}
\text{Label}(x_i) = 
\begin{cases} 
0 & \text{if } \|x_i - \mu_1\| \leq \|x_i - \mu_2\| \\
1 & \text{otherwise}
\end{cases}
\end{equation}
where 0 is the benign case and 1 is the anomalous case.

\section{Results and Evaluation}


\subsection{Datasets}


For the network-only source domain dataset, we utilize WUSTL-2021 \cite{wustl-2021-paper}. This dataset was collected from an Industrial IoT (IIoT) testbed designed to monitor the water level and turbidity quantity of a water storage tank. 93\% of the dataset is benign data and 7\% of the dataset contains four attack types: denial-of-service (89.98\%), reconnaissance (9.46\%), SQL command injection (0.31\%), and backdoor (0.25\%). There are 49 initial features, but in line with the original work \cite{wustl-2021-paper}, we utilized the identified subset of 23 intrusion-detection specific features. 

We utilize the Robot Operating System 2 for Smart Passenger Center (ROSPaCe) dataset \cite{Puccetti2024ROSPaCe} as our CPS-domain dataset. ROSPaCe is an intrusion dataset collected from the Smart Passenger Center (SPaCe) embedded cyber-physical system. The SPaCe system is used to orchestrate and monitor public mobility by surveying train cars and determining their current operative conditions, utilizing key perception components included in a variety of critical CPS. The entire dataset contains 30,247,050 data points and 481 columns. 25 of these features come from the OS monitor, 5 come from the ROS Monitor Node, and 452 from Tshark, a command-line network traffic analysis tool. ROSPaCe includes data collected from the operating system, ROS, and network layers of the SPaCe architecture. The dataset is imbalanced, with 23 million attack data points (78\%) and over 6.5 million normal data points (22\%). In our work, we utilize the lightweight version of the dataset. It contains the same number of data points and only removes feature data, keeping the 60 best-performing features alongside the label and timestamp columns. In line with the original ROSPaCe work, we also utilize these 60 features. ROSPaCe contains six different attacks, two of which are discovery attacks and four of which are DoS attacks. 

More details about the two datasets are summarized in Table \ref{tab:dataset_descriptions}.

\subsection{Clustering-based AD Results}

We utilize PyOD \cite{zhao2019pyod}, a Python toolbox that provides benchmark anomaly detection models for multivariate datasets, to establish baseline accuracy metrics for our datasets. We utilize the Deep Support Vector Data Description (DSVDD) \cite{deepsvdd}, Gaussian Mixture Models (GMM) \cite{GMM_source}, and lightweight on-line detector of anomalies (LODA) methods provided by PyOD for both WUSTL-2021 and ROSPaCe. All baselines used are unsupervised models where labels are not included during model fitting. Our baseline models can be trained in two ways: (1) using only benign data and tested on unseen benign and anomaly samples, or (2) using a mixed subset of benign and anomaly data for training and tested on a separate unseen set. We include both sets of results here for reference in Table \ref{tab:baselinespyod_and_ours}. We note here that the PyOD baselines are trained on individual flows and not the sequenced flows created for our proposed methodology. 


We also include our proposed method's results in Table \ref{tab:baselinespyod_and_ours} for both the source and target domains, utilizing our clustering-based anomaly decision method. While the source domain data (WUSTL-2021) utilizes the label classifier during training, in evaluation, we utilize the proposed clustering-based anomaly decision methods on the source domain after training as well. This allows us to evaluate if the anomaly decision method presented can be used for our new environment and for unseen data from the original source domain after training. In the final version of the model, the latent dimension is fixed at 512, the sequence length for both domains is fixed at 25, and the loss weights are: $\lambda_{Dunn}$ = 1, $ \lambda_{TML}$ = 0.1, $\lambda_{class}$, and $\lambda_{D}$ = 1.



We find that the PyOD models tend to perform better when fitted with benign data and tested on a mixture of unseen benign and anomaly data, particularly with the ROSPaCe dataset. For instance, for ROSPaCe, we observe a 74\% overall accuracy increase between the mixed and benign LODA models. Regarding the performance of our proposed method, we find that when using our proposed anomaly decision method with unseen source domain data after training, we are able to outperform the baselines in overall accuracy, precision and F-one scores while performing comparably with the baselines with recall. With the ROSPaCe target domain data, we find that our method performs comparably with the benign-only trained baselines from PyOD and greatly outperforms the mixed training set baselines from PyOD. An advantage of our method compared to PyOD baselines lies in the use of sequenced system metric flows as we can aggregate metrics and perform periodic detection to save system resources.

\section{Conclusion}

Attack detection in CPS can be challenging due to their multi-layer architectures, imbalanced datasets, and lack of labeled data. Here, we present a novel CPS anomaly detection  method aimed at bridging the gap between the strength of network-only IDS systems and the multi-layer architecture of most CPS. Our method allows us to transfer the attack information from a labeled, network-only source domain to a unlabeled, CPS target domain. We show through accuracy metrics that our method can perform comparably to current outlier detection models while using aggregated network flows formulated as sequences. This work is a key contribution towards closing the gap in comprehensive, multi-layer CPS-targeted IDS.

\bibliography{references}
\bibliographystyle{ieeetr}

\vspace{12pt}

\end{document}